\documentclass{article}
\usepackage{spconf,amsmath,amssymb,graphicx,subcaption,hyperref}

\title{INVESTIGATING LABEL NOISE SENSITIVITY OF CONVOLUTIONAL NEURAL NETWORKS FOR FINE GRAINED AUDIO SIGNAL LABELLING}
\name{Rainer Kelz \qquad Gerhard Widmer} 
\address{
  Johannes Kepler University Linz\\
  Department of Computational Perception\\
  Altenberger Str. 69, 4040 Linz, Austria
}

\def\fps{{\,[\mathrm{fps}]}}
\newcommand*\dif{\mathop{}\!\textnormal{\slshape d}}

\def\fround{{\mathrm{round}}}
\def\fceil{{\mathrm{ceil}}}
\def\ffloor{{\mathrm{floor}}}

\begin{document}
%
\maketitle
\begin{abstract}
  We measure the effect of small amounts of systematic and random label noise caused by slightly misaligned ground truth labels in a fine grained audio signal labeling task. The task we choose to demonstrate these effects on is also known as \textit{framewise polyphonic transcription} or \textit{note quantized multi-f0 estimation}, and transforms a monaural audio signal into a sequence of note indicator labels. It will be shown that even slight misalignments have clearly apparent effects, demonstrating a great sensitivity of convolutional neural networks to label noise. The implications are clear: when using convolutional neural networks for fine grained audio signal labeling tasks, great care has to be taken to ensure that the annotations have precise timing, and are free from systematic or random error as much as possible - even small misalignments will have a noticeable impact.
\end{abstract}
\begin{keywords}
convolutional neural networks, multi-label classification, framewise polyphonic transcription
\end{keywords}
\section{Introduction}
\label{sec:intro}
Recent empirical work on quantifying the generalization capabilities of deep neural networks \cite{Arpit_2017, Zhang_2016} questions the usefulness of traditional learning theory applied to neural networks, and among other things shows that image classification performance gradually degrades in the presence of mislabeled images.

We investigate how severe these effects are for time series labeling when the ground truth labels are only slightly misaligned, which is a common occurrence when dealing with manually annotated time series data. The assumption being that the main difference between label noise in time series and label noise for images caused by annotators is the similarity of examples in input space. This comparison is justified by the fact that the way a sequence labeling task with convolutional neural networks is usually set up, corresponds exactly to repeated image classification on similar images obtained by shifting a reading window across the short time Fourier transformed audio signal. However, because the distribution of examples in the image domain is different from examples obtained in the audio domain it is not immediately clear to which extent label noise is a problem.

\begin{figure}[htb]
\begin{minipage}[b]{1.0\linewidth}
  \centering
  \centerline{\includegraphics[width=7.5cm]{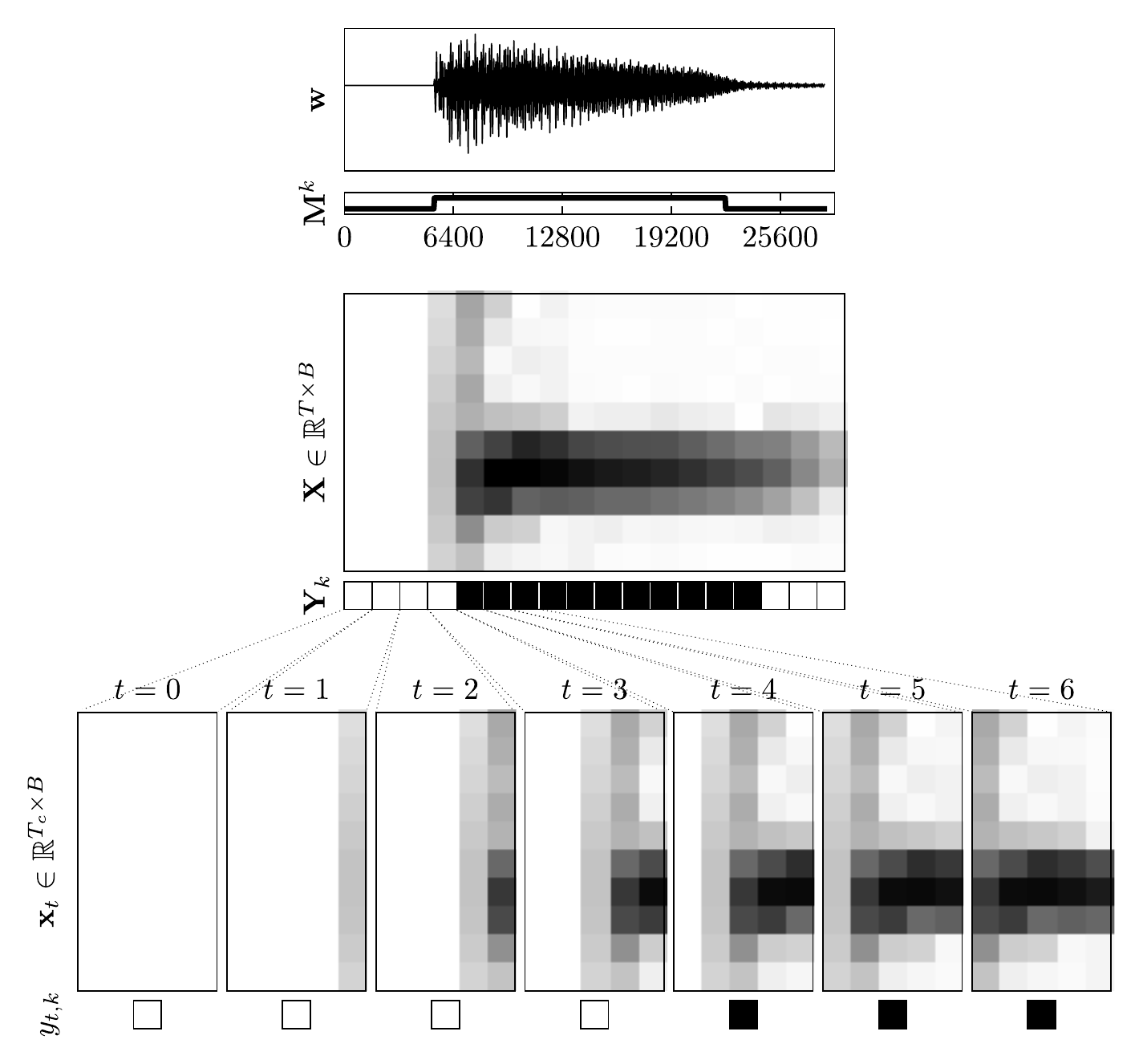}}
\end{minipage}
\caption{The intuitive reason why imprecision in audio signal annotation may yield highly similar, yet differently labeled examples. At the last stage of the signal processing chain we see pairs of input and corresponding indicator $(\mathbf{x}_t, y_{t,k})$ for label $k$. Note the very similar frames $\mathbf{x}_3$ and $\mathbf{x}_4$ and their different labels.}
\label{fig:labelnoise_schema}
\end{figure}

We posit that it is highly unlikely that two similar images will be assigned different labels by the same annotator. It is more likely that an imprecision either in hand movements steering a pointing device such as the mouse, or the annotation software used itself, will lead to very similar examples in time being assigned different labels. A sketch of this intuitive notion can be seen in  figure \ref{fig:labelnoise_schema}, where we can observe the sequential transformations of the audio signal input together with its annotation.

The data consists of pairs $\mathbf{w} \in \mathbb{R}^{T^{hi}}$ denoting the audio signal and $\mathbf{M} \in \{0, 1\}^{T^{hi} \times K}$ denoting the annotation, where $T^{hi}$ is the number of audio samples, $K$ is the number of labels, and $k$ is the index of an arbitrary label. A common preprocessing step is the short-time Fourier transform (STFT) of $\mathbf{w}$ and subsequent application of a filter bank to obtain $\mathbf{X} \in \mathbb{R}^{T^{lo} \times B}$. $B$ is the frequency resolution, dependent on the choice of filter bank applied after the Fourier transform. The usually much lower frame rate of the STFT is dependent on the hop size, resulting in $T^{lo} \ll T^{hi}$. In a similar fashion, the high resolution annotation $\mathbf{M}$ is transformed into $\mathbf{Y} \in \mathbb{R}^{T^{lo} \times K}$ to match up with the filtered STFT.

The final inputs to the convolutional neural network are pairs $\mathbf{x}_t \in \mathbb{R}^{T_c^{lo} \times B}$ of excerpts of length $T_c^{lo}$ from $\mathbf{X}$ at time $t$ and labels $y_{t, k} \in \{0, 1\}$ as a target for each label indicator output. Looking at these steps in detail makes it apparent that slight misalignments in the annotation, be they random or systematic, lead to collections of pairs $\{(\mathbf{x}_a, y_{a,k}), (\mathbf{x}_b, y_{b,k}), \dots\}$ where most distance measures in input space $d(\mathbf{x}_a, \mathbf{x}_b)$ are small but the targets for these examples differ, as $y_{a,k} \neq y_{b,k}$.

The sequence labeling task we chose to investigate the impact of misaligned annotations on, is also called \textit{framewise polyphonic transcription} or \textit{note quantized multi-f0 estimation} in the music information retrieval community. We claim that the effects measured here also extend to beyond the framewise scenario because the misalignment problems we consider only affect the start and end positions of a label, and hence also extend to systems that try and predict labels in interval form.

\section{Models}
\label{sec:models}
The convolutional neural networks we use for time series labeling are parametrized functions $g_{\theta}: \mathbb{R}^{T_c \times B} \rightarrow \{0,1\}^{K}$, mapping excerpts from a filtered STFT of length $T_c$ in time and width $B$ in frequency to a vector of length $K$, whose components indicate the presence or absence of a label. After the application of a logarithmic filter bank to the STFT, the number of bins comes down to $B=229$, as described in \cite{Kelz_2016}. For framewise transcription of pianos, $K=88$ denotes the tonal range of the instrument, and a label indicator having a value of $1$ means that a note is sounding in the excerpt presented as the input. The misalignment effects are measured at two different STFT frame rates, $31.25 \fps$ and $100 \fps$, the lower frame rate also being used in \cite{Sigtia_2016, Kelz_2016}. Keeping the temporal context approximately the same for the two frame rates necessitated the use of two different architectures, with the one for the higher frame rate being deeper and wider, yet only slightly increasing parameter count. \footnote{Source code to replicate all results can be found at \url{https://github.com/rainerkelz/ICASSP18}}

For the training procedure, we adhere closely to the description in \cite{Kelz_2016}, which uses mini-batch stochastic gradient descent with \mbox{Nesterov} momentum and a step-wise learning rate schedule, but reduced mini-batch size. A rather drastic reduction of the number of examples in a batch from $128$ examples as advocated in \cite{Kelz_2016}, to $8$ examples in the present work, is motivated by findings in \cite{Keskar_2016}, which state that noisier gradient estimates are helpful in finding flatter minima, thus potentially improving generalization. We notice a small improvement in prediction performance together with a convenient reduction in training time.

\section{Dataset}
\label{sec:dataset}
We chose the MAPS dataset \cite{Emiya_2010} as our experimental testbed due to the availability of a very precise ground truth, free from any human annotator disagreement. Note that for the purpose of demonstrating non-negligable effect sizes of misalignments, any annotation could suffice in principle, as long as it is unambiguous. The data consists of a collection of MIDI files, and corresponding audio renderings. Multiple sample banks were used to render the audio files. To further increase acoustic variability, several MIDI files were played back on a computer controlled Disklavier and recorded in close and ambient microphone conditions. The MIDI files in the MAPS dataset have a sufficiently high temporal resolution that enables us to neglect any quantization error stemming from the conversion of MIDI ticks to seconds and treat the start and end times of note labels effectively as if they were originating from a continuous space.
There are two train-test protocols defined in \cite{Sigtia_2016}, with different amounts of instrument overlap. In \textit{Configuration-I}, instruments in training, validation and test sets overlap, whereas in \textit{Configuration-II} only training and validation sets contain overlapping instruments. The test set for \textit{Configuration-II} solely consists of pieces rendered with the Disklavier. For both configurations, the $31.25 \fps$ models are trained and evaluated on four different splits of the training data. The models with the higher frame rate input at $100 \fps$ are trained and evaluated on one fold only for both configurations due to the high computational cost.

\section{Experimental Setup}
\label{sec:experimental_setup}
For all models trained, all non-architectural hyper parameters are fixed, the only varied quantities are the choice of frame rate and the choice of labeling function, the exact notion of which we will now define.

\begin{figure}
\begin{subfigure}{0.5\textwidth}
\begin{center}
  \includegraphics[width=8.5cm]{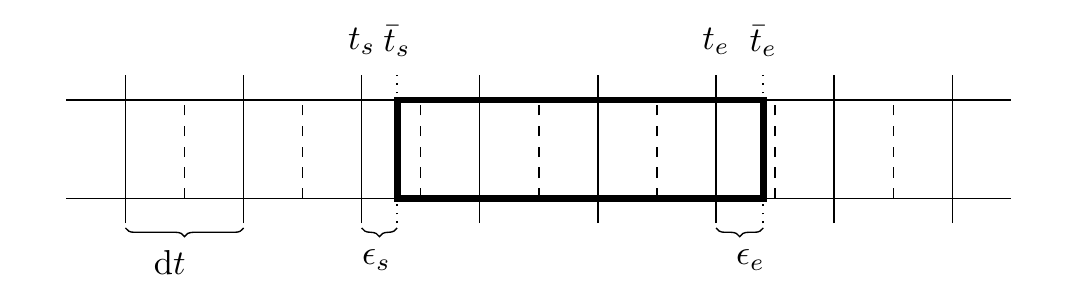}
  \caption{schematic illustration of quantities involved in the definition of labeling functions for fine grained sequence labeling tasks}
  \label{fig:legend}
\end{center}
\end{subfigure}
\begin{subfigure}{0.5\textwidth}
\begin{center}
\begin{tabular}{c|l|l}
  $f_{\cdot}(\bar{t}_s, \bar{t}_e)$  & \multicolumn{1}{c|}{$t_s$}            & $t_e$ \\
\hline
$f_a$    & $\lfloor \bar{t}_s / \dif t \rceil$   & $\lfloor \bar{t}_e / \dif t \rceil$ \\

  $f_b$    & $\lceil \bar{t}_s / \dif t \rceil$    & $\lceil \bar{t}_e / \dif t \rceil$ \\
  $f_c$    & $\lfloor \bar{t}_s / \dif t \rfloor$  & $\lfloor \bar{t}_e / \dif t \rfloor$ \\
$f_d$    & $\lfloor \bar{t}_s / \dif t \rfloor$  & $\lfloor \bar{t}_e / \dif t \rfloor + \lfloor (\bar{t}_e - \bar{t}_s) / \dif t \rfloor$ \\
\hline
  $f_e$    & $f_a + R_j$                       & $f_a + R_j$ \\
  $f_f$    & $f_a + R_s$                       & $f_a + R_e$
\end{tabular}
\end{center}
\caption{The different labeling functions used, leading to different kinds of quantization error. Symbols $\lfloor \cdot \rfloor, \lfloor \cdot \rceil, \lceil \cdot \rceil$ denote the functions $\ffloor(\cdot), \fround(\cdot), \fceil(\cdot)$ respectively, and $R_{j, s, e} \in \{ -1, 0, 1 \}$ are discrete, uniformly distributed random variables. $f_a$ is used as the reference labeling function throughout this article.}
\label{tbl:labellingfunctions}

\end{subfigure}
\caption{Quantization schemes and definitions of labeling functions.}
\end{figure}

Throughout this paper we will use the term \textit{labeling function} to mean the complete conversion process from high resolution annotations to assigning concrete labels \mbox{$y_{t,k} \in \{0,1\}$} to the examples $\mathbf{x}_t \in \mathbb{R}^{T_c \times B}$ shown to the network. This includes the annotator as well, be it a mechanistic generator, as in the case of extracting labels from MIDI files, or a human providing manual annotation. The main focus for each function lies on the conversion from high resolution annotations to lower resolution annotations. Figure \ref{fig:legend} shows a schematic depiction of the conversion of a label from an interval defined with high time resolution into a sequence of much fewer labels at a lower time resolution. Symbols $\bar{t}_{s}, \bar{t}_{e}$ denote start and end times in high resolution, expressed in seconds (we pretend these are continuous, due to their high time resolution). Symbols $t_{s}, t_{e}$ are their counterparts in low resolution, expressed as discrete frame indices, with $\epsilon_{s}, \epsilon_{e}$ denoting the errors incurred by rounding after conversion. The symbol $\dif t$ denotes the length of one lower resolution frame in seconds and is used as the conversion factor.

The exact definitions of the different labeling functions used to transform the high resolution annotations obtained from MIDI files into framewise labels can be found in figure \ref{tbl:labellingfunctions}. They can all be viewed as functions of the form $f: \mathbb{R} \times \mathbb{R} \rightarrow \mathbb{N} \times \mathbb{N}$, mapping pairs of continuous times to pairs of discrete indices.

The differences between labeling functions lie in the choice of how to quantize and when. Functions $f_{\{a,b,c,d\}}$ deal with systematic misalignment caused by systematic quantization errors, and functions $f_{\{e,f\}}$ randomly modify the result of $f_{a}$, by either shifting both start and end indices jointly by a random variable $R_j \in \{-1, 0, 1\}$, or shifting the two indices separately by two independent random variables $R_s, R_e \in \{-1, 0, 1\}$. For all functions involving random variables, their realizations are drawn from a discrete uniform distribution at the time of conversion. This means that for each pair of start and end times and a particular experimental run, a potential shift can happen only once.

We treat the labeling function $f_a$, which simply rounds to the nearest integer, as the reference. All evaluations are done on the first $30 [\mathrm{s}]$ of all pieces in the respective test set for a fold, and against a ground truth obtained at a frame rate of $100 \fps$ with labeling function $f_a$. This means the predictions of the models running at a lower frame rate of $31.25 \fps$ need to be upsampled again for evaluation. To measure prediction performance, we use the definitions for precision $\mathcal{P}$, recall $\mathcal{R}$ and f-measure $\mathcal{F}$ as described in \cite{Bay_2009}. 

\newcommand{\sumT}{\sum_{t=1}^{T}}

\begin{center}
\begin{align}
\mathcal{P} & = \frac{\sumT TP[t]}{\sumT TP[t] + FP[t]} \\
\mathcal{R} & = \frac{\sumT TP[t]}{\sumT TP[t] + FN[t]} \\
\mathcal{F} & = \frac{2 \cdot \mathcal{P} \cdot \mathcal{R}}{\mathcal{P} + \mathcal{R}}
\end{align}
\end{center}

\section{Results}
\label{sec:results}
\begin{figure}[htb]
\begin{minipage}[b]{1.0\linewidth}
  \centering
  \centerline{\includegraphics[width=8cm]{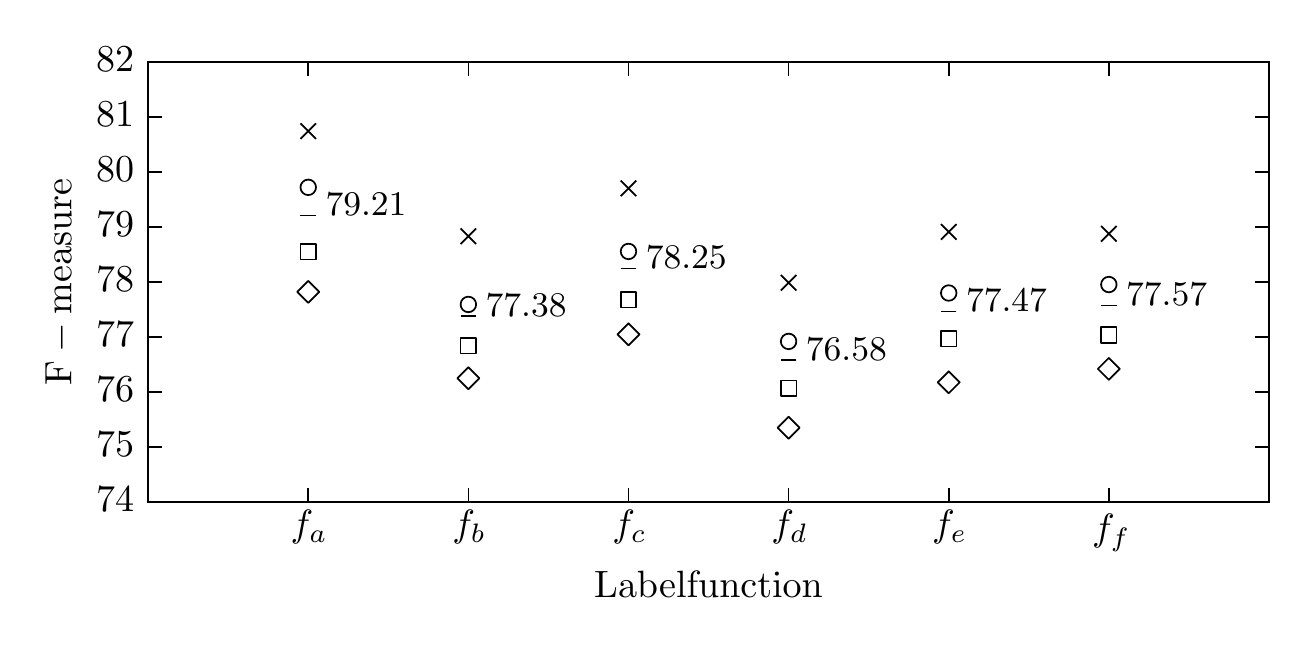}}
\end{minipage}
\caption{Configuration I: F-measure on the validation set at a frame rate of $31.25 \fps$ across different labeling functions for multiple folds.}
\label{fig:conf_1_fps_31_25_valid_summary_fmeasure}
\end{figure}

\begin{figure}[htb]
\begin{minipage}[b]{1.0\linewidth}
  \centering
  \centerline{\includegraphics[width=8cm]{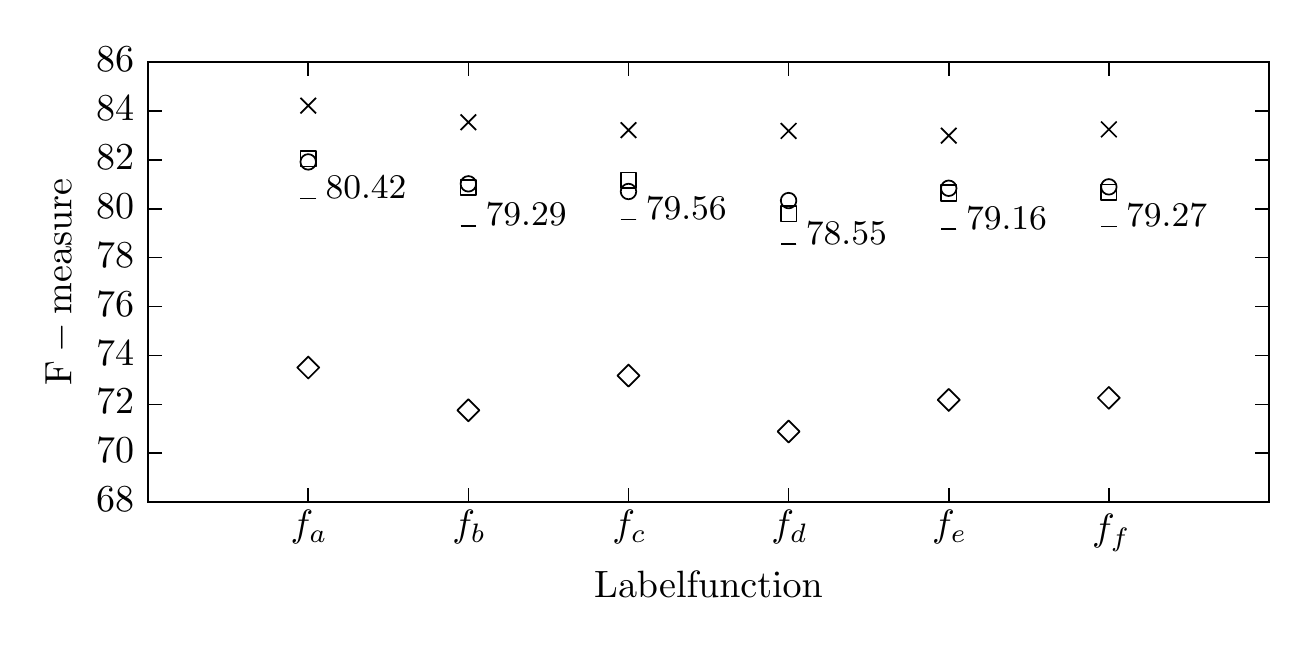}}
\end{minipage}
\caption{Configuration I: F-measure on the test set at a frame rate of $31.25 \fps$ across different labeling functions for multiple folds.}
\label{fig:conf_1_fps_31_25_test_fmeasure}
\end{figure}

The effect of small misalignments on label annotation can be observed in figures \ref{fig:conf_1_fps_31_25_valid_summary_fmeasure} and \ref{fig:conf_1_fps_31_25_test_fmeasure}, which show clear evidence of the effect of using different labeling functions at a frame rate of $31.25 \fps$, across multiple folds for Configuration I with similar data in train and test sets. Along the horizontal axis we see the labeling functions, and the vertical axis shows f-measure.

For each labeling function we see the performance on individual folds ($\times, \circ, \square, \diamond$) and the textually annotated mean ($-$) of all folds. We immediately notice that a labeling function with either a small systematic ($f_{\{a, b, c, d\}}$) or random ($f_{\{e, f\}}$) error has a non-negligible effect on the f-measure.

The impact on the performance on the test set is less severe, as can be observed in figure \ref{fig:conf_1_fps_31_25_test_fmeasure}, but still on the order of $1$ percentage point. Incidentally, the mean result for the reference labeling function improves slightly on the state of the art for this task as reported in \cite{Kelz_2016}.

\begin{figure}[htb]
\begin{minipage}[b]{1.0\linewidth}
  \centering
  \centerline{\includegraphics[width=8cm]{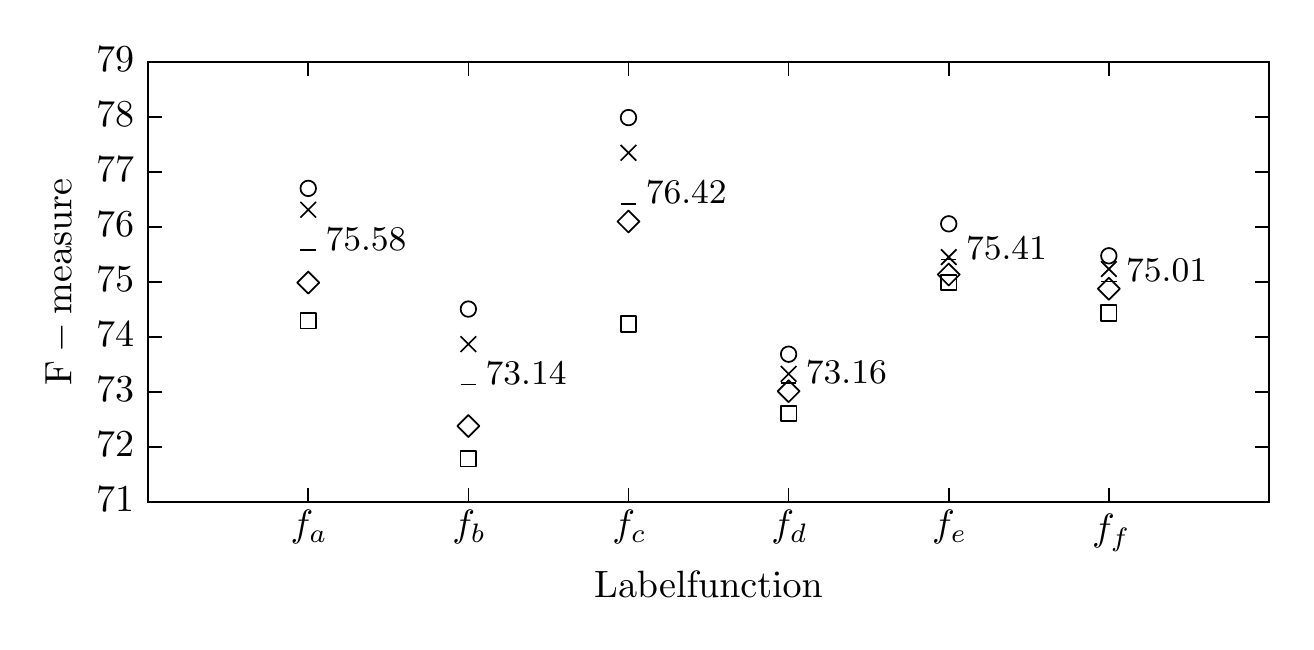}}
\end{minipage}
\caption{Configuration II: F-measure on the test set at a frame rate of $31.25 \fps$ across different labeling functions for multiple folds.}
\label{fig:conf_2_fps_31_25_test_fmeasure_boxplot}
\end{figure}

Interestingly, when inspecting the lower frame rate results for Configuration II which has much more dissimilar train and test sets in terms of acoustic conditions, we can still observe a similar pattern. Even small misalignments lead to non negligible differences in performance, observable on the test set results in figure \ref{fig:conf_2_fps_31_25_test_fmeasure_boxplot}. The obtained results all improve upon the state of the art as reported in \cite{Kelz_2016}, regardless of labeling function which we attribute to some extent to the much smaller batch size and the usage of a labeling function introducing systematic error in \cite{Kelz_2016}.


\begin{figure}[htb]
\begin{minipage}[b]{1.0\linewidth}
  \centering
  \centerline{\includegraphics[width=8cm]{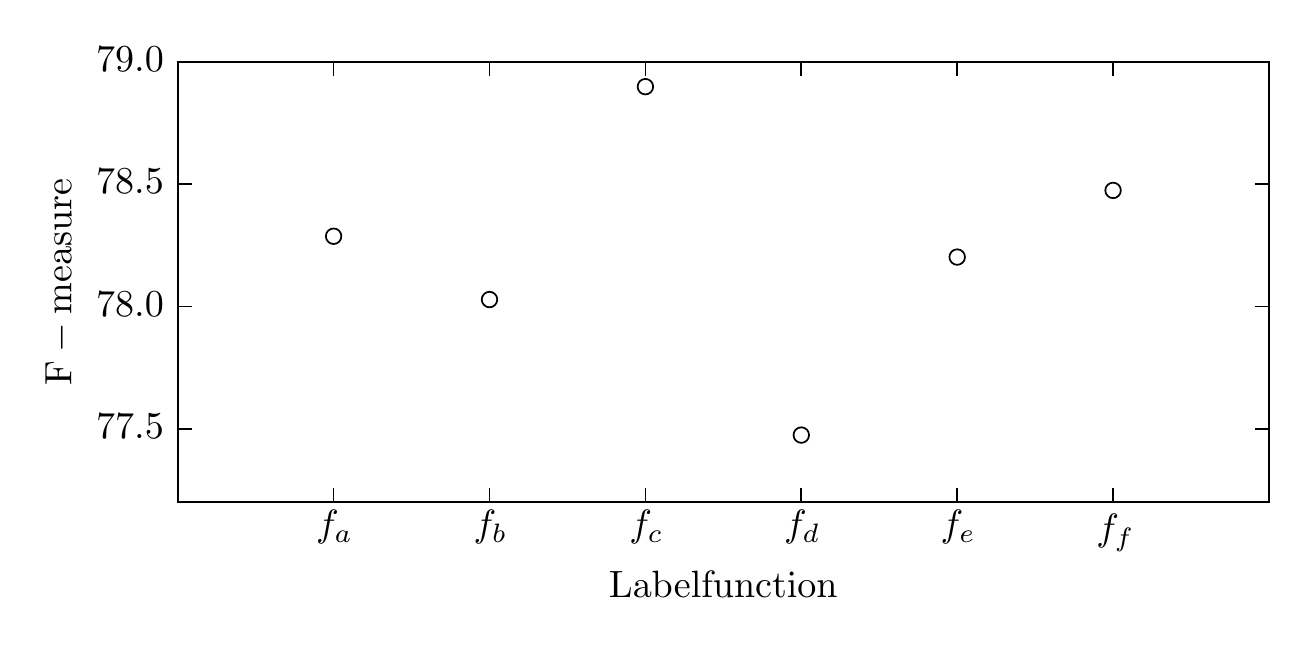}}
\end{minipage}
\caption{Configuration II: F-measure on the test set at a frame rate of $100 \fps$ across different labeling functions.}
\label{fig:conf_2_fps_100_test_fmeasure_boxplot}
\end{figure}

Finally, focusing our attention on the results for the test set at a higher frame rate of $100 \fps$ in figure \ref{fig:conf_2_fps_100_test_fmeasure_boxplot}, a pattern of performance differences with respect to the reference labeling function $f_a$ is still noticeable. The severity is much smaller however, which is to be expected, due to the higher frame rate and hence lower quantization error. Results are shown for only one fold, due to the high computational costs of training and evaluation at higher frame rates.

An interesting oddity in both low and high frame rate cases for Configuration II, is the performance increase when using $f_c$ to train and $f_a$ to obtain the ground truth for evaluation. This indicates a problem with the ground truth alignment, likely due to MIDI clock drift or similar issues, and will need to be addressed in future work. It has no effect on the main contribution of this work, which is to demonstrate that convolutional neural networks are highly sensitive to even small amounts of label noise, and to increase awareness that this issue needs to be addressed for fine grained \mbox{audio} signal labeling tasks.




\section{Conclusion}
The effect of systematic and random label noise stemming from small misalignments of label annotations on convolutional neural networks in the context of audio signal labeling was investigated empirically, and shown to be non-negligible. We therefore conclude that great care must be taken to make sure the ground truth annotations align with the events in the audio as much as possible, especially if the intention is to use the data for fine grained sequence labeling, for example for subsequent analysis of musical timing. We demonstrated the effect with the help of already very precise annotations stemming from a mechanistic generator. We surmise that even in the case of having multiple annotations of human origin, and hence the potential to use annotator disagreement to pinpoint problematic labels and so obtain better estimates of the true label alignment, the sensitivity issue will need to be addressed carefully.

\section{Acknowledgments}
A Tesla K40 used for parts of this research was donated by the NVIDIA Corporation. We also thank the authors and maintainers of the Lasagne \cite{lasagne} and Theano \cite{theano} software packages.


\bibliographystyle{IEEEbib}
\bibliography{master}

\end{document}